\documentclass[twocolumn,showpacs,preprintnumbers,amsmath,amssymb,final]{revtex4}

\usepackage{graphicx}
\usepackage{dcolumn}
\usepackage{bm}

\def\BEq{\begin{equation}}
\def\EEq{\end{equation}}
\def\BEqA{\begin{eqnarray}}
\def\EEqA{\end{eqnarray}}
\def\BEn{\begin{enumerate}}
\def\EEn{\end{enumerate}}
\def\BWT{\begin{widetext}}
\def\EWT{\end{widetext}}
\def\a{\alpha}

\def\e{\epsilon}

\def\w{\omega}

\def\bra{\langle}

\def\ket{\rangle}

\begin{document}


\title{Single-step implementation of the
controlled-Z gate in a qubit/bus/qubit device}

\author{Andrei Galiautdinov}
 \affiliation{Department of Electrical
Engineering, University of California, Riverside, California 92521,
USA}

\date{\today}

\begin{abstract}

We propose a simple scheme for generating a high-fidelity 
controlled-Z (CZ) 
gate in a three-component qubit/bus/qubit device.
The corresponding tune/detune pulse is {\it single-step}, with a near-resonant constant undershoot between the $|200\ket$ and $|101\ket$ states. During the pulse, the frequency of the first qubit is kept fixed, while
the frequency of the second qubit is varied in such a way as to bring 
the $|200\ket$ and $|101\ket$ states close to resonance.
As a result, the phase of the $|101\ket$ state is accumulated via the corresponding {\it second-order} anticrossing. For experimentally realistic qubit frequencies and a 75 MHz coupling (150 MHz splitting), 
a 45 ns gate time can be realized with $>99.99$\% intrinsic fidelity, with
errors arising due to the non-adiabaticity of the ramps. 
The CZ pulse is characterized by {\it two} adjustable parameters: the undershoot magnitude and undershoot duration. The pulse does {\it not} load an excitation into the bus. This by-passes the previously proposed
need for two additional qubit-to-bus and bus-to-qubit MOVE operations. Combined with the recently predicted high-fidelity idling operation in the RezQu architecture [A. Galiautdinov, J. Martinis, A. Korotkov (unpublished)], this controlled-Z scheme may prove useful for implementations on the first generation quantum computers.

\end{abstract}


\maketitle


\section{introduction}

Recent progress in preparing, controlling, and measuring the macroscopic
quantum states of superconducting circuits with Josephson junctions
\cite{YAMAMOTO2003,STEFFEN2006,PLANTENBERG2007,NEELEY2008,
DICARLO2009,BIALCZAK2010,REED2010} makes realization of a quantum computer
an experimental possibility \cite{CLARKE2008}. 
Two major roadblocks --
decoherence and scalability -- may soon be overcome by the so-called
Resonator/zero-Qubit (RezQu) architecture, recently proposed by 
J. Martinis \cite{RezQu}. Some of the basic operations of the RezQu 
architecture (such as the idling operation, the generation and measurement 
of the single-excitation states, as well as the single-excitation transfer 
operation called MOVE) were analyzed in a joint paper \cite{IDLING-PAPER}. 
It was found, that the RezQu architecture 
is capable of providing high-fidelity performance required for 
quantum information processing. 

In spite of the optimistic conclusions presented in Ref. \cite{IDLING-PAPER},
an important problem of generating {\it high-fidelity entangling}
operations in the RezQu architecture still remains. 
One such operation is the controlled-Z (CZ) gate, given in Eq. (\ref{eq:CZmatrix}).
It is believed that, in the RezQu architecture, the CZ gate may easily be 
produced using the SWAP-based {\it three-step} approach, similar to that of 
Ref. \cite{Haack-2010}, in which
one logic qubit is moved to the bus, transferring the qubit excitation 
onto the bus, while the other qubit is tuned close to resonance with the bus
for a precise duration. After the needed phase is accumulated 
(as in Refs. \cite{Strauch-2003,Yamamoto-10}), the excitation is moved back from 
the bus to the original qubit. We have simulated this three-step approach 
for realistic RezQu parameters and found some difficulties with it, which are 
described in Section \ref{sec:problems}. This prompted us to look for a more 
direct scheme, which is not beset by such difficulties. The scheme is described 
in Sec. \ref{sec:Implementation}. It does not rely on the loading and unloading 
of the bus, but instead, uses a second order anticrossing for the required phase 
accumulation ({\it cf.} Ref. \cite{Zheng-2009}). 

\section{The qubit/bus/qubit device}

\begin{figure}
\includegraphics[angle=0,width=1.00\linewidth]{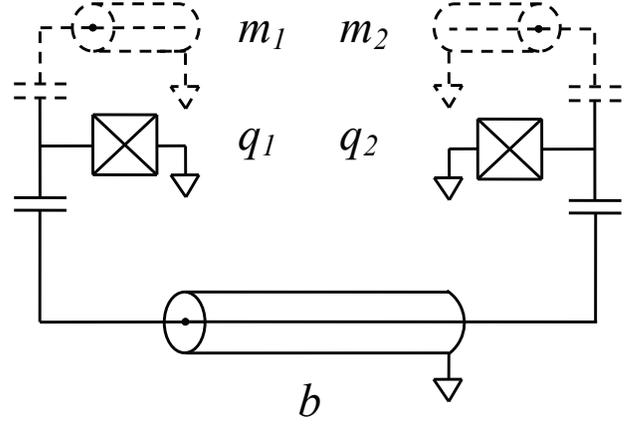}
\caption{ \label{fig:1} Schematic diagram of a qubit/bus/qubit RezQu device. 
The qubits may be supplemented with memory resonators (dashed lines). 
Here, $q$ -- qubits, $b$ -- bus, $m$ -- memory resonators.}
\end{figure}

A three-component RezQu device is depicted in Fig. \ref{fig:1}.
In the rotating wave approximation (RWA), its dynamics is described by the Hamiltonian
\BEqA
\label{RWAhamiltonian1}
&&H(t) = 
\sum_{i=1,2}H_{i}(t)
+ \w_b a^{\dagger}_b a_b 
\nonumber \\
&&
+ g_{b1} 
\left(\sigma_{1}^{-}a^{\dagger}_b + \sigma_{1}^{+}a_b\right)
+ g_{b2}\left(a^{\dagger}_b \sigma_{2}^{-} + a_b\sigma_{2}^{+} \right),
\EEqA
where
\BEq
H_{i}(t) = 
\begin{bmatrix}
0 & 0 & 0\cr
0 & \w_{i}(t) &0 \cr
0&0& 2\w_{i}(t) - \eta_{i}
\end{bmatrix}
\EEq
are the Hamiltonians of the qubits whose frequencies $\w_{i}$ 
may vary in time and 
the anharmonicities $\eta_i$ are assumed to be constant, 
\BEq
\sigma_{i}^{-} = 
\begin{bmatrix}
0 & 1 &0\cr
0 & 0 &\sqrt{2} \cr
0 & 0 &0
\end{bmatrix},
\quad
\sigma_{i}^{+} = \left(\sigma_{i}^{-}\right)^{\dagger},
\EEq
are the qubit lowering and raising operators, $\w_b$ is 
the bus frequency (which is held fixed), 
$a^{\dagger}_b$ and $a_b$ are the creation 
and annihilation operators 
for the bus photons, and $g_{b1}$, $g_{b2}$ are the bus-qubit 
coupling constants. In our numerical simulations we will assume
that $\eta_1=\eta_2\equiv \eta$ and $g_{b1}=g_{b2}\equiv g_b$.

\section{Some difficulties with the 
SWAP-based controlled-Z gate implementation}

\label{sec:problems}

\begingroup
\begin{table}
\caption{
\label{tab:1}
Configuration of the $(q_1 b q_2)$ system before, after, and during 
the CZ gate. All frequencies are defined by $\nu = \omega/2\pi$ (GHz) 
and are {\it bare}. Both qubit anharmonicities are $\eta/2\pi = 0.2$ GHz 
and assumed to be constant. The coupling is chosen to be $g_b/2\pi = 75$ 
MHz to guarantee sufficiently fast, $t_{\rm gate} = 45$ ns, 
gate operation. Left arrows indicate the near-resonant two-excitation
frequencies.}
\begin{ruledtabular}
\begin{tabular}{lll}
		Bare						& Before and & Optimized CZ frequencies at \\
		frequencies						& after CZ	& the $200\leftrightarrow 101$ 
																	anticrossing\\
								\hline
   $\nu_{q1}$&6.6&6.6\\
   $\nu_{b}$&6.0&6.0\\
   $\nu_{q2}$&6.5&6.40959\\
   $\nu_{110}$&12.6&12.6\\
   $\nu_{101}$&13.1&13.00959 $\leftarrow$\\
   $\nu_{011}$&12.5&12.40959\\
   $\nu_{200}$&13.0&13.0 $\leftarrow$\\
   $\nu_{020}$&12.0&12.0\\
   $\nu_{002}$&12.8&12.61918\\
   
 \end{tabular}
\end{ruledtabular}
\end{table}
\endgroup

\begin{figure}
\includegraphics[angle=0,width=1.00\linewidth]{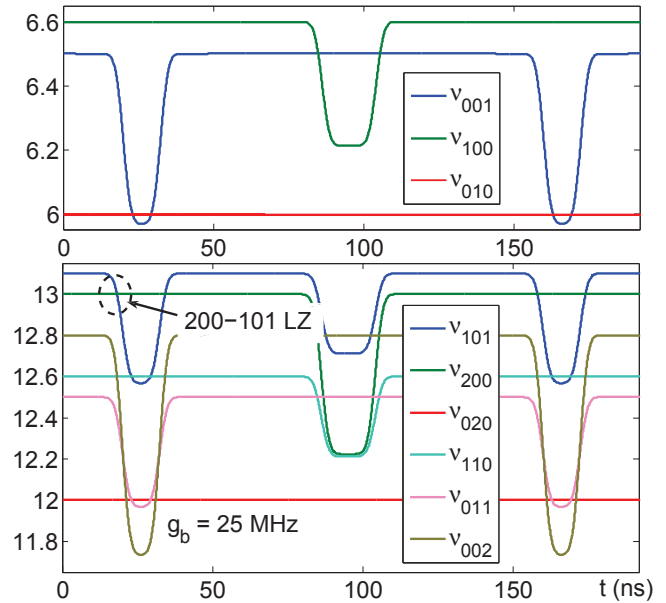}
\caption{ \label{fig:2} 
(Color online) One- and two-excitation frequencies (in GHz)
of the qubit/bus/qubit system in the usual 
3-step SWAP-based CZ gate implementation. 
Optimization with at least {\it four} naturally chosen parameters 
at $g_b/2\pi = 25$ MHz gives the 
gate error, as defined in Eq. (\ref{eq:gateAccuracy}), 
of no less than $2\times 10^{-3}$. 
Compare with Fig. \ref{fig:3}, where an alternative, single-step
CZ-generating scheme is presented. }
\end{figure}

In what follows, we assume that the qubit/bus/qubit $(q_1 b q_2)$ system 
starts and ends 
in the (default) configuration, as shown in Table \ref{tab:1}. The  
initial and final qubit frequencies, $\w_{q1}>\w_{q2}$, are
chosen in such a way as to avoid the $|100\ket \leftrightarrow |001\ket$ 
and $|200\ket \leftrightarrow |101\ket$ crossings. 
Then, the standard \cite{Yamamoto-10} 
three-step SWAP-based CZ gate implementation 
(Fig. \ref{fig:2})
suffers from the following major drawback: it produces a large number of 
Landau-Zener (LZ) transitions, each of which degrades the resulting gate's 
fidelity. The RezQu architecture based on fixed 
couplings may not be flexible enough to provide the needed controllability 
to counter the effects of all these transitions in a three-step CZ manner. 
Controlling only the qubit frequencies may not be enough to achieve the needed 
$10^{-4}$ accuracy of the CZ gate, using {\it experimentally reasonable number} 
of parameters. As Fig. \ref{fig:2} shows, the very first ramp of the initial 
SWAP operation already contains one such LZ crossing, leading to the leakage 
from state $|101\ket$ to state $|200\ket$. 

\begin{figure}
\includegraphics[angle=0,width=1.00\linewidth]{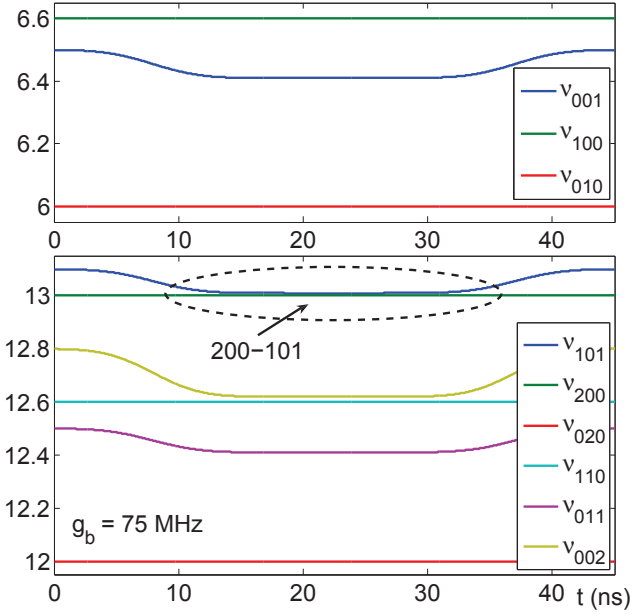}
\caption{ \label{fig:3} 
(Color online) One- and two-excitation frequencies (in GHz)
of the qubit/bus/qubit system in the single-step CZ gate 
implementation. A two-parameter optimization at $g_b/2\pi=75$ MHz
gives $>99.99$\% fidelity for total gate duration of $t_{\rm gate} = 45$ ns. 
The optimized parameters are the undershoot,
$(\omega_{101}-\omega_{200})/2\pi=9.59$ MHz, and the undershoot duration,
$t_{\rm undershoot}=29.1$ ns (measured between the central 
points of the two ramps). 
The widths of the error-function-shaped ramps were held fixed at 
$\sigma_{\rm in}=\sigma_{\rm fin}=3$ ns. 
Compare with Fig. \ref{fig:2}.}
\end{figure}

Here we propose to turn this particular drawback into an asset by 
dropping the loading and unloading SWAP operations altogether and 
employing the mentioned 
$|200\ket \leftrightarrow |101\ket$ anticrossing to accumulate 
the needed 101-phase during the CZ operation (see Fig. \ref{fig:3}).


\section{Potential problems with the proposed scheme}

The following two problems may arise in our scheme. 

First, the presence
of additional system elements (qubits, memory resonators, etc.) may 
result in additional 
states that are near-resonant with the states $|200\dots\ket$ and 
$|101\dots\ket$, thus leading to unwanted leakage. Here we ignore 
this complication and assume that under realistic conditions it will 
always be possible to isolate this particular anticrossing sufficiently 
well.

Second, being a second-order process, accumulation of the 101-phase 
may 
proceed too slowly compared with the qubit coherence time (currently at 
about $t_{\rm coherence} \simeq 500$ ns). However, 
the following argument shows that this is not necessarily true. 

In the case of a similar second order resonance 
$|100\ket \leftrightarrow |001\ket$, the effective 
(via the bus) $q_1$-$q_2$ 
coupling \cite{Pinto-2010} is given by
\BEq
g_{\rm eff}^{100\leftrightarrow 001} = \frac{2g_b^2\w_b}{\w_{q1}^2-\w_b^2}.
\EEq
Then in our case we should have
\BEq
g_{\rm eff}^{200\leftrightarrow 101} 
= \sqrt{2}g_{\rm eff}^{100\leftrightarrow 001}.
\EEq
Choosing $g_b=75$ MHz (experimentally achievable coupling)
and setting $\w_{q1}/2\pi \approx 6.4$ GHz (near-resonant condition), 
we find for $\w_{b}/2\pi = 6.0$ GHz,
\BEqA
\frac{g_{\rm eff}^{200\leftrightarrow 101}}{2\pi} 
&=&\sqrt{2} \left(\frac{2\times 0.075^2\times 6.0}{6.4^2-6.0^2}\right)
\nonumber \\
&\approx& 0.0192 \; {\rm GHz} = 19.2 \; {\rm MHz},
\EEqA
which gives an experimentally reasonable duration of
the corresponding phase accumulation,
\BEq
t_{2\pi{\rm \; pulse}}^{200\leftrightarrow 101} 
= \frac{\pi}{g_{\rm eff}^{200\leftrightarrow 101}} 
\approx 26 \; {\rm ns}.
\EEq
In the actual implementation shown in Fig. \ref{fig:2}, the total gate duration
had to be prolonged to $t_{\rm gate} = 45$ ns in order to correctly produce 
the final populations of $|100\ket$ and $|001\ket$ states.

\section{Implementing the single-step CZ gate}

\label{sec:Implementation}

\begin{figure}
\includegraphics[angle=0,width=1.00\linewidth]{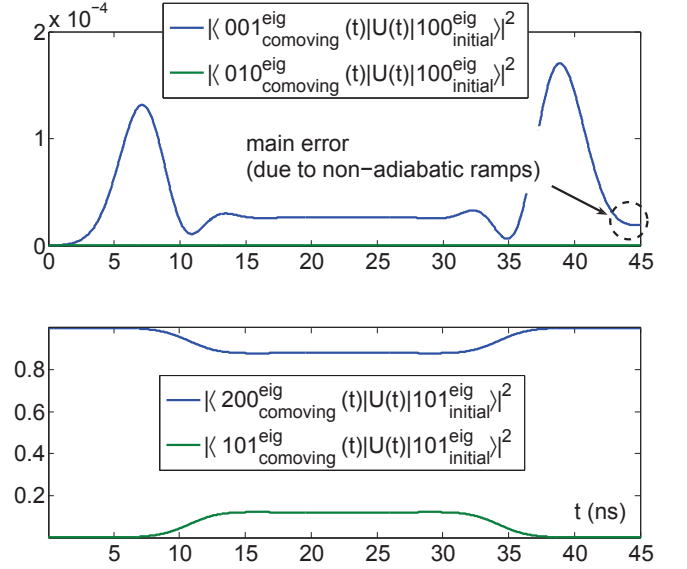}
\caption{ \label{fig:4} 
(Color online) Some of the overlaps between 
the time-evolving computational
states and the comoving system eigenstates in the single-step 
CZ gate implementation.
$U(t)$ stands for the unitary operator of the time evolution up to time $t$.}
\end{figure}

The proposed single-step CZ gate resulting from a two-parameter 
optimization with the RWA Hamiltonian of Eq. (\ref{RWAhamiltonian1})
is depicted in Fig. \ref{fig:3}.
To provide some intuitive understanding of how the generated gate 
works, the overlaps between the time-evolved logic states and some 
of the time-dependent (comoving) system eigenstates are given in 
Fig. \ref{fig:4}. 

Our gate is implemented in the computational basis consisting of 
the full system eigenstates \cite{IDLING-PAPER}. It has the form
\BEq
\label{eq:CZmatrix}
{\rm CZ} =
\begin{pmatrix}
1&0&0&0\cr
0&e^{-i\varphi_1}&0&0\cr
0&0&e^{-i\varphi_2}&0\cr
0&0&0&-e^{-i(\varphi_1+\varphi_2)}\cr
\end{pmatrix},
\EEq
where $\varphi_1$ and $\varphi_2$ are some arbitrarily 
accumulated phases.
Due to the use of the system {\it eigenstates}, these phases 
can always be adjusted simply by waiting.
The optimization was performed at fixed $t_{\rm gate} = 45$ ns 
by minimizing the function 
\BEqA
\label{eq:gateAccuracy}
{\rm Error}(U)
&=& {\rm Error}_1+{\rm Error}_2+{\rm Error}_3+{\rm Error}_4
\nonumber \\
&=&\left(1-|a_1|^2\right) 
+ \left(1-|a_2|^2\right) 
+ \left(1-|a_3|^2\right) 
\nonumber \\
&&
+ 
\left\vert 
1
+\frac{
a_1a_2a^*_3}{\left\vert a_1a_2a^{*}_3 \right\vert}
\right\vert
 \geq 0,
\EEqA
where
\BEq
a_1 =
\bra \overline{100} |U| \overline{100}\ket,
a_2 =
\bra \overline{001} |U| \overline{001}\ket,
a_3 =
\bra \overline{101} |U| \overline{101}\ket,
\EEq
with respect to the undershoot magnitude and the undershoot duration 
(measured between the central 
points of the ramps), with additional constraints
${\rm Error}_1 +{\rm Error}_2+{\rm Error}_3<10^{-4}$ 
and ${\rm Error}_4<10^{-10}$. 
The widths (standard deviations) of the error-function-shaped ramps
were held fixed at 3 ns.
The results are presented in Fig. \ref{fig:3}.
In the above, $U$ is the unitary operator representing the 
CZ pulse, and the overbars stand for the prefix ``eigen-.''
The optimization function ${\rm Error}(U)$ was defined so 
that for $U={\rm CZ}$, as given in Eq. (\ref{eq:CZmatrix}), 
${\rm Error}({\rm CZ})=0$. Notice, that we do not have to take into account 
the phase of the $|000\ket$ state, since the corresponding frequency 
$\nu_{000}$ can always be set to 0. 

\section{CZ gate as an idling error}

Our CZ gate may be viewed as a particular example of an ``idling error'' 
\cite{IDLING-PAPER}, which is a measure of how
fast the phase of the {\it computational} eigenstatestate $|\overline{101}\ket$ accumulates
relative to the phases of eigenstates $|\overline{100}\ket$ and $|\overline{001}\ket$. 
The error is characterized by the running frequency
$ \Omega_{ZZ} = \varepsilon_{101}-\varepsilon_{100}-\varepsilon_{001}+\varepsilon_{000}$,
with $\varepsilon_{ijk}$ being the corresponding
eigenenergies, and physically arises due to the level repulsion between 
101 and other levels in the two-excitation subspace of the system. 
Consequently, a superposition of computational states evolves as
\BEqA
|\psi(t)\ket
&=&\a_{000} e^{-i\e_{000}t} |\overline{000}\ket 
+ \a_{100} e^{-i\e_{100}t} |\overline{100}\ket 
\nonumber \\
&&
 + \a_{001} e^{-i\e_{001}t} |\overline{001}\ket
\nonumber \\
&&
+ \a_{101}e^{-i\Omega_{ZZ}t}
e^{-i(\e_{100}+\e_{001}-\e_{000})t}|\overline{101}\ket,
\EEqA
and so, after a time $t_{\rm cp} =\pi/\Omega_{ZZ}$,
the state $|\overline{101}\ket$ gets multiplied by -1. 
Thus, in systems with nonlinearities, the CZ gate can always be 
generated simply by waiting.
For the qubit/bus/qubit RezQu device, using Eq. (\ref{RWAhamiltonian1}), 
we find in fourth order,
\BWT
\BEq
\label{eq:OmegaZZ_4th_order}
\Omega^{(4)}_{ZZ} =
\frac{
2g_{b1}^2 g_{b2}^{2}
\left\{
\w_1 \eta_1 (2\w_b-\w_1-\eta_2) + \w_2 \eta_2(2\w_b-\w_2-\eta_1)
- \w_b\left[\w_b (\eta_1+\eta_2)-2\eta_1\eta_2\right]
\right\}
}
{(\w_1-\w_b)^2(\w_2-\w_b)^2
\left[\w_1-(\w_2-\eta_2)\right]\left[(\w_1-\eta_1)-\w_2\right]},
\EEq
\EWT
which for the above mentioned (and fixed) $\w_1/2\pi = 6.6$ GHz and 
$\w_2/2\pi = 6.5$ GHz, at $g_b/2\pi = 75$ coupling, produces the CZ gate 
after about 130 ns,
which is too long. 
For a more efficient CZ generation, the system parameters must be set to 
maximize $\Omega_{ZZ}$. One such choice, $\w_2 \approx \w_1-\eta_1$, which 
corresponds to the $200-101$ anticrossing, was made in our single-step implementation.

\section{Conclusion}

To summarize, we introduced a scheme for single-step generation of 
a high-fidelity
controlled-Z gate in a three-component RezQu architecture. 
Despite the use of a second-order anticrossing, the accumulation of the 
needed 101-phase proceed sufficiently fast compared with the qubit coherence 
time. Different from the usually considered proposals, 
our CZ scheme does not 
rely on the MOVE operations transferring excitations to and from the bus. 
The resulting simplicity of the generated gate may prove useful for 
implementations in the first generation solid state quantum computers.

\begin{acknowledgments}
This work was supported by NSA/IARPA/ARO Grant No. W911NF-10-1-0334.

\end{acknowledgments}


\begin{thebibliography}{}


\bibitem{YAMAMOTO2003}
Y. Yamamoto et al, Nature {\bf 425}, 941 (2003).

\bibitem{STEFFEN2006}
M. Steffen et al., Science {\bf 313}, 1423 (2006).

\bibitem{PLANTENBERG2007}
J. H. Plantenberg et al., Nature {\bf 447}, 836 (2007).

\bibitem{NEELEY2008}
M. Neeley et al., Nature Physics {\bf 4}, 523 (2008).

\bibitem{DICARLO2009}
L. DiCarlo et al, Nature {\bf 460}, 240 (2009).

\bibitem{BIALCZAK2010}
R. C. Bialczak et al., Nature Physics {\bf 6}, 409 (2010).

\bibitem{REED2010}
M. D. Reed et al., Phys. Rev. Lett. 105, 173601 (2010).

\bibitem{CLARKE2008}
J. Clarke and F. K. Wilhelm, Nature {\bf 453}, 1031 (2008).

\bibitem{RezQu}
J. M. Martinis (unpublished).

\bibitem{IDLING-PAPER}
A. Galiautdinov, J. Martinis, A. Korotkov, presented at the 
2011 APS March Meeting, Dallas, TX (unpublished).

\bibitem{Pinto-2010}
R. A. Pinto et al., Phys. Rev. B {\bf 82}, 104522 (2010).

\bibitem{Haack-2010}
G. Haack et al., Phys. Rev. B {\bf 82}, 024514 (2010).

\bibitem{Strauch-2003}
F. W. Strauch et al., Phys. Rev. Lett. {\bf 91}, 167005 (2003).

\bibitem{Yamamoto-10} T. Yamamoto et al., Phys. Rev. B {\bf 82},
184515 (2010).

\bibitem{Zheng-2009}
Shi-Biao Zheng, Appl. Phys. Lett. {\bf 94}, 154101 (2009).


\end{thebibliography}
\end{document}